\documentclass[aps,prl,preprint,showpacs,showkeys,floatfix,nofootinbib,superscriptaddress,a4paper]{revtex4}

\newcommand{\bmath}[1]{\mbox{\boldmath $#1$}}
\usepackage{graphicx,amssymb,amsmath,color}

\usepackage{dcolumn}
\usepackage{bm}

\begin{document}
\title{Precision measurements of the \bmath{pp\to \pi^+pn}
and \bmath{pp\to \pi^+d} reactions: importance of long-range and
tensor force effects}
%
\author{A.~Budzanowski}
\affiliation{Institute of Nuclear Physics, PAN, Krakow, Poland}
\author{A.~Chatterjee}
\affiliation{Nuclear Physics Division, BARC, Mumbai, India}
\author{P.~Hawranek}
\affiliation{Institute of Physics, Jagellonian University, Krakow,
Poland}
\author{R.~Jahn}
\affiliation{Helmholtz-Institut f\"{u}r Strahlen- und Kernphysik der
Universit\"{a}t Bonn, Bonn, Germany}
\author{V.~Jha}
\affiliation{Nuclear Physics Division, BARC, Mumbai, India}
\author{K.~Kilian}
\affiliation{Institut f\"{u}r Kernphysik, Forschungszentrum J\"{u}lich,
J\"{u}lich, Germany}
\affiliation{J\"{u}lich Centre for Hadron Physics, Forschungszentrum J\"{u}lich, J\"{u}lich, Germany}
\author{Da.~Kirillov}
\affiliation{Institut f\"{u}r Kernphysik, Forschungszentrum J\"{u}lich,
J\"{u}lich, Germany}%
\affiliation{Fachbereich Physik, Universit\"{a}t Duisburg-Essen,
Duisburg, Germany}
\affiliation{J\"{u}lich Centre for Hadron Physics, Forschungszentrum J\"{u}lich, J\"{u}lich, Germany}
\author{Di.~Kirillov}
\affiliation{Laboratory for High Energies, JINR Dubna, Russia}
\author{S.~Kliczewski}
\affiliation{Institute of Nuclear Physics, PAN, Krakow, Poland}
\author{D.~Kolev}
\affiliation{Physics Faculty, University of Sofia, Sofia, Bulgaria}
\author{M.~Kravcikova}
\affiliation{Technical University, Kosice, Kosice, Slovakia}
\author{M.~Lesiak}
\affiliation{Institut f\"{u}r Kernphysik, Forschungszentrum J\"{u}lich,
J\"{u}lich, Germany}%
\affiliation{Institute of Physics, Jagellonian
University, Krakow, Poland}
\affiliation{J\"{u}lich Centre for Hadron Physics, Forschungszentrum J\"{u}lich, J\"{u}lich, Germany}
\author{J.~Lieb}
\affiliation{Physics Department, George Mason University, Fairfax,
Virginia, USA}
\author{H.~Machner}
\email{h.machner@fz-juelich.de}%
\affiliation{Institut f\"{u}r Kernphysik, Forschungszentrum
J\"{u}lich, J\"{u}lich, Germany}%
\affiliation{Fachbereich Physik, Universit\"{a}t Duisburg-Essen,
Duisburg, Germany}
\affiliation{J\"{u}lich Centre for Hadron Physics, Forschungszentrum J\"{u}lich, J\"{u}lich, Germany}
\author{A.~Magiera}
\affiliation{Institute of Physics, Jagellonian University, Krakow,
Poland}
\author{R.~Maier}
\affiliation{Institut f\"{u}r Kernphysik, Forschungszentrum J\"{u}lich,
J\"{u}lich, Germany}
\affiliation{J\"{u}lich Centre for Hadron Physics, Forschungszentrum J\"{u}lich, J\"{u}lich, Germany}
\author{G.~Martinska}
\affiliation{P.~J.~Safarik University, Kosice, Slovakia}
\author{S.~Nedev}
\affiliation{University of Chemical Technology and Metalurgy, Sofia,
Bulgaria}
\author{J.~A.~Niskanen}
\affiliation{Department of Physics, University of Helsinki, Finland}
\author{N.~Piskunov}
\affiliation{Laboratory for High Energies, JINR Dubna, Russia}
\author{D.~Proti\'c}
\affiliation{Institut f\"{u}r Kernphysik, Forschungszentrum J\"{u}lich,
J\"{u}lich, Germany}
\affiliation{J\"{u}lich Centre for Hadron Physics, Forschungszentrum J\"{u}lich, J\"{u}lich, Germany}
\author{J.~Ritman}
\affiliation{Institut f\"{u}r Kernphysik, Forschungszentrum J\"{u}lich,
J\"{u}lich, Germany}
\affiliation{J\"{u}lich Centre for Hadron Physics, Forschungszentrum J\"{u}lich, J\"{u}lich, Germany}
\author{P.~von~Rossen}
\affiliation{Institut f\"{u}r Kernphysik, Forschungszentrum J\"{u}lich,
J\"{u}lich, Germany}
\affiliation{J\"{u}lich Centre for Hadron Physics, Forschungszentrum J\"{u}lich, J\"{u}lich, Germany}
\author{B.~J.~Roy}
\affiliation{Nuclear Physics Division, BARC, Mumbai, India}
\author{I.~Sitnik}
\affiliation{Laboratory for High Energies, JINR Dubna, Russia}
\author{R.~Siudak}
\affiliation{Institute of Nuclear Physics, PAN, Krakow, Poland}
\author{H.~J.~Stein}
\affiliation{Institut f\"{u}r Kernphysik, Forschungszentrum J\"{u}lich,
J\"{u}lich, Germany}
\affiliation{J\"{u}lich Centre for Hadron Physics, Forschungszentrum J\"{u}lich, J\"{u}lich, Germany}
\author{R.~Tsenov}
\affiliation{Physics Faculty, University of Sofia, Sofia, Bulgaria}
\author{J.~Urban}
\affiliation{P.~J.~Safarik University, Kosice, Slovakia}
\author{G.~Vankova}
\affiliation{Institut f\"{u}r Kernphysik, Forschungszentrum J\"{u}lich,
J\"{u}lich, Germany}%
\affiliation{Physics Faculty, University of Sofia,
Sofia, Bulgaria}
\affiliation{J\"{u}lich Centre for Hadron Physics, Forschungszentrum J\"{u}lich, J\"{u}lich, Germany}
\author{C.~Wilkin}
\affiliation{Department of Physics and Astronomy, UCL, London, U.K.}
\collaboration{The COSY-GEM Collaboration} \noaffiliation
\date{\today}
%
%
\begin{abstract}%

Inclusive measurements of pion production in proton--proton
collisions in the forward direction were undertaken at 400 and
600~MeV at COSY using the Big Karl spectrograph. The high resolution
in the $\pi^+$ momentum ensured that there was an unambiguous
separation of the $pp\to {\pi}^+d/\pi^+pn$ channels. Using these and
earlier data, the ratio of the production cross sections could be
followed through the $\Delta$ region and compared with the
predictions of final state interaction theory. Deviations are
strongly influenced by long-range terms in the production operator
and the tensor force in the final $pn$ system. These have been
investigated in a realistic $pp\to\pi^+d/\pi^+pn$ calculation that
includes $S \rightleftharpoons D$ channel coupling between the final
nucleons. A semi-quantitative understanding of the observed effects
is achieved.
\end{abstract}

\maketitle
%
%

Pion production in nucleon-nucleon collisions at intermediate
energies involves a delicate interplay between the basic production
mechanism and the strong interactions between the two or three
particles in the final state. Information on this might be obtained
by comparing the $pp\to \pi^+d$ and $pp\to \pi^+pn$ reactions. With
this in mind, we measured simultaneously the two final states for
forward-going pions at a proton beam energy of $T_p=951$~MeV by
studying the inclusive $pp\to \pi^+X$ reaction and achieving a mass
resolution of $\sigma_X =97$~keV/$c^2$ in the region of the deuteron
peak~\cite{Abdel-Bary05a}. This is almost four times better than that
of the previous best experiment~\cite{Pleydon99} and means that
events corresponding to the $\pi^+d$ two-body final state could be
unambiguously separated from those of the $\pi^+pn$ continuum.
Moreover, the high resolution was sufficient to show that the
production of $S$-wave spin-singlet $\{pn\}_s$ final states is
negligible compared to the spin-triplet $\{pn\}_t$.

Simultaneous measurement of the $\pi^+d$ and $\pi^+pn$ final states
allows one to evaluate the cross section ratio
\begin{equation}\label{equ:Ratio}
R_{pn/d}=\left.\frac{d^2\sigma}{d\Omega\,d{x}}
(pp\to\pi^+\left\{pn\right\}_t)\right/ \frac{d\sigma}{d\Omega}(pp\to
\pi^+d)
\end{equation}
with few systematic errors, being untroubled by questions of relative
normalization of different experiments. Here $x=\varepsilon/B$ is the
excitation energy $\varepsilon$ in the $pn$ system in units of the
deuteron binding energy $B$.

If the coupling between the $S$ and $D$ states through the tensor
force is neglected, then the F\"{a}ldt-Wilkin final state interaction
(FSI) theorem~\cite{FW96} shows that
\begin{equation}\label{equ:d_pn}
R_{pn/d}\approx N\,\frac{p(x)}{p(-1)}
\frac{\sqrt{x}}{2\pi(x+1)}\:,
\end{equation}
where $p(x)$ and $p(-1)$ are the pion c.m.\ momenta for the $\pi^+pn$
and $\pi^+d$ channels, respectively. The normalization factor $N$,
which must be unity at the deuteron pole when $x=-1$, should differ
little from this above threshold provided that the pion production
operator is of short range~\cite{FW96}. However, a value of $N$ close
to two was required in order to fit the 951~MeV
results~\cite{Abdel-Bary05a}.

The effects of the tensor force have been studied for pion production
in $pp\to\pi^+d$, where dramatic changes were found when the
$S\rightleftharpoons D$ coupling was
introduced~\cite{Niskanen78,Niskanen94}. Since, on kinematic grounds,
the $D$-state amplitudes might be expected to change sign between the
deuteron bound state and the $pn$ continuum, it was
suggested~\cite{Abdel-Bary05a} that the discrepancy in the value of
$N$ might be explained as being due to $S$--$D$ interference that was
not included in the FSI theorem~\cite{FW96}. However, no detailed
theoretical estimates had been made of the influence of the tensor
force for the $\pi^+\left\{pn\right\}_t$ continuum final state and it
is not clear whether it is this or the finite range of the production
operator that is the crucial feature.

The 951~MeV data were taken above the peak associated with the
production of the $\Delta$ resonance. Since the $S$--$D$ interference
is predicted to change sign around the
resonance~\cite{Abdel-Bary05a,Niskanen78, Niskanen94}, we have
repeated the experiment at one energy below the $\Delta$ peak
($T_p\approx 400$~MeV) and at another close to it ($T_p\approx
600~$MeV).

The present experiment is a straightforward extension of that
previously reported~\cite{Abdel-Bary05a} and so only the salient
points of the set-up will be described here. Since we are interested
principally in the ratio of the cross sections for the $pp\to
\pi^+d/\pi^+\{pn\}$ reactions as a function of the excitation energy
$\varepsilon$ of the $pn$ system, it is crucial to measure this
energy with high resolution for well identified pions. In order to
optimize the momentum resolution of the system, the proton beam of
the COSY synchrotron was electron cooled at injection, accelerated to
the requested momentum, and then stochastically extracted. The
resulting beam was focussed onto the center of a target cell that was
only 2~mm thick, with windows made of 1~$\mu$m Mylar. The beam spot
was less than 0.5~mm in diameter, with divergences of 1.1 and
1.3~mrad in directions perpendicular to the beam. These
characteristics were much better than those achieved without beam
cooling and, in particular, the background was considerably reduced.

Positive pions were detected near the forward direction using the Big
Karl magnetic spectrograph~\cite{Drochner98}. Their positions and
track directions in the focal plane were measured with two sets of
multiwire drift chambers, each composed of six layers. The chambers
were followed by scintillator hodoscopes that were used to measure
the time of flight over a distance of 3.5~m, which led to the
unambiguous $\pi^+$ identification. The set-up allowed the excitation
energy $\varepsilon$ in the $pn$ system to be determined with high
precision.

\begin{figure}[!h]
\begin{center}
\includegraphics[width=0.4\textwidth]{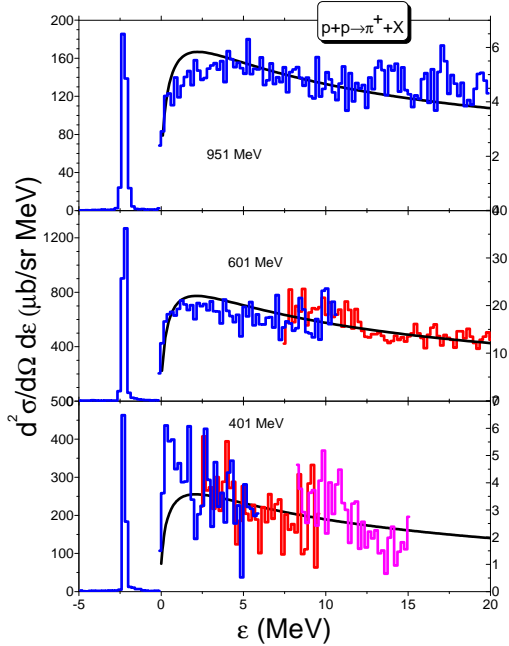}
\caption{(Color online) Forward differential cross sections for the
$pp\to \pi^+X$ reaction for three different beam energies as
functions of the $pn$ excitation energy $\varepsilon$. The left axis
is for $X=d$ while the right are for the $X=pn$ continuum. The
measurements are shown as histograms and the colors indicate
different settings of the magnetic spectrograph. The curves are the
results of the F\"{a}ldt-Wilkin model of Eq.~(\ref{equ:Ratio}), with
the normalization factors $N$ being given in Table~\ref{Tab:Norm}. }
\label{fig:spectra}
\end{center}
\end{figure}

Count rates were converted to cross sections by normalizing the
integrated deuteron peaks to the zero-degree $pp\to \pi^+d$ values,
for which there are extensive data~\cite{SAIDpid}. The results of
these and our previous measurement~\cite{Abdel-Bary05a} are shown in
Fig.~\ref{fig:spectra} as a function of $\varepsilon$. Although
corrections for acceptance etc., are included, these are slowly
varying for $\varepsilon < 20$~MeV. It is clear that there is an
excellent separation between the $\pi^+\{pn\}$ and $\pi^+d$ channels,
which leads to robust determinations of $R_{pn/d}$. There is also no
sign of any $\{pn\}_{\!s}$ spin-singlet production, which would show
up as a narrow peak for $\varepsilon \approx 1$~MeV. An analysis of
the isospin-related $pp\to \{pp\}_{\!s}\pi^0$
reaction~\cite{Betigeri02} shows that the cross section should be
less than 0.1~$\mu$b/sr\,MeV.
%
%
%
%

Also shown in Fig.~\ref{fig:spectra} are the $pp\to\pi^+\{pn\}$
predictions of the FSI theorem~\cite{FW96}, with the normalization
factors $N$ required to bring the calculations and data into
agreement for $\varepsilon\approx 10$~MeV being given in
Table~\ref{Tab:Norm}. The values of $N$ depend somewhat on the
interval considered as well as on the background assumptions. Similar to Ref. \cite{Pleydon99} we subtracted a constant background extending from the deuteron peak to the continuum, which is 1.5, 4.8, and 0.66~$\mu$b/(sr\,MeV) and were included in the curves shown at 401, 601 and 951 MeV, respectively.

%
%
\begin{table}[htb]
\caption{Normalization factor $N$ of Eq.~(\ref{equ:d_pn}) for various
beam energies as determined by fitting data at $\varepsilon\approx
10$~MeV. The results of Ref.~\cite{Pleydon99} required an
extrapolation in angle to compare with the GEM results in the forward
direction.}\label{Tab:Norm}
\begin{ruledtabular}
\begin{tabular}{ccc}
$T_p$ & Normalization & Reference        \\ %
(MeV) & factor $N$    &                  \\ \hline %
401  & $0.51\pm0.06$ & This work        \\
420  & $0.94\pm0.07$ & \cite{Pleydon99} \\
500  & $1.11\pm0.07$ & \cite{Pleydon99} \\
601  & $1.06\pm0.04$ & This work        \\
951  & $2.2\pm0.1$ & \cite{Abdel-Bary05a}    \\
\end{tabular}
\end{ruledtabular}
\end{table}
%
%

Our measurements can be supplemented by the results obtained at
TRIUMF at 420 and 500~MeV~\cite{Pleydon99}. Although the resolution
in $\varepsilon$ was somewhat poorer than that achieved with Big
Karl, it was sufficient to separate the $\pi^+d$ and $\pi^+pn$ final
states. The data, which typically cover a range from $24^{\circ}$ to
$100^{\circ}$ in laboratory angle, were transformed into the c.m.\
system to allow values of $R_{pn/d}$ to be extracted, assuming that
the two cross sections were to be evaluated for the same beam energy
and production angle $\theta_{\pi}$. The normalization factors are
also given in Table~\ref{Tab:Norm}, where some of the error arises
from that in the angular extrapolation to the forward direction.

In order to study the results in more detail, the previous two-body
calculations~\cite{Abdel-Bary05a} have been extended to the full
three-body final state. In the presence of the tensor force, this
involves the nontrivial task of evaluating slowly converging overlap
integrals of $S$ and $D$ final scattering wave functions with the
initial $pp$ states. As in Ref.~\cite{Niskanen92}, these were
performed by rotating the contour integration in the region outside
the range of the $pn$ potential into the positive or negative
imaginary directions, as required to ensure convergence.  The
numerical evaluations were carried out with the Reid soft core
potential~\cite{Reid68}, which is quite adequate at low energies.

In addition to the direct production from distorted $NN$ states,
$s$-wave pion rescattering was also taken into account. If half of
the available energy in the $NN$ case is ascribed to the intermediate
pion then this can place it on the energy shell for beam energies
above the two-pion threshold~\cite{Koltun66, Niskanen78}. This
therefore also leads to a long range mechanism and an integral which
has to be handled in the same way as that for the direct production
amplitude. Rescattering in the $p$ wave through the $\Delta(1232)$
isobar is accounted for by the $\Delta N$ admixtures generated in the
initial states by the coupled channels method (with associated
changes in the $NN$ force). As discussed for the case of a decaying
$\Delta$ in Ref.~\cite{Niskanen78}, all the final pion energy is
assumed to be already in the intermediate pion, putting it on the
energy shell. However, in this case the $\Delta N$ wave function is
of finite range and this raises no convergence issues.

%
%
\begin{figure}[hbt]
\begin{center}
\includegraphics[width=0.4\textwidth]{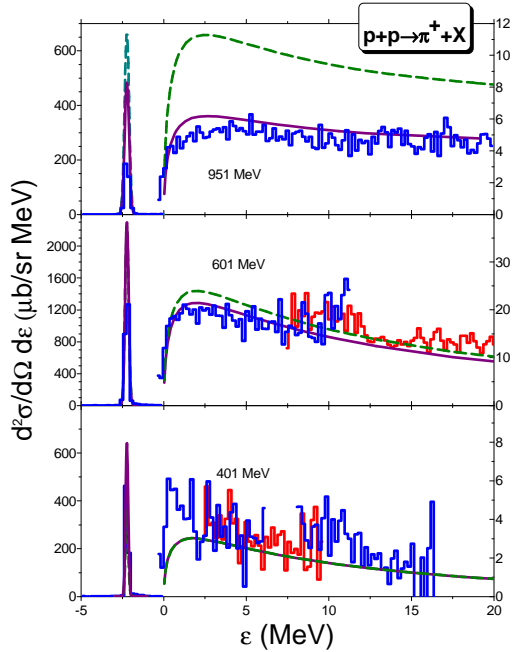}
\caption{(Color online) Same as Fig.~\ref{fig:spectra} but for two-
and three-body final state calculations. Solid curves include $S$-$D$
interference, which is omitted for the dashed curves.}
\label{Fig:Data-Jouni}
\end{center}
\end{figure}
%
%

In order to investigate in detail the influence of the $D$-state
component, the full tensor coupling with mixed $S+D$ final state was
first studied, with the results of the model calculations being
compared with the data in Fig.~\ref{Fig:Data-Jouni}. The tensor force
was then switched off and the intermediate range $pn$ attraction
multiplied by a factor of 1.3 to reproduce the correct deuteron
binding energy. The results are plotted with a resolution determined
by a Gaussian fit to the deuteron peak and a constant background
included, as for Fig.~\ref{fig:spectra}. The calculations with the
tensor force account well for the data in the continuum, though they
overestimate the deuteron contribution at the highest beam energy.
The calculations without $S$-$D$ interference also lie well above the
data here, whereas the agreement with the data is very good for the
two lower energies. The two types of calculations almost agree with
one another at 400 and 600~MeV.

%
%
\begin{figure}[!h]
\begin{center}
\includegraphics[width=0.4\textwidth]{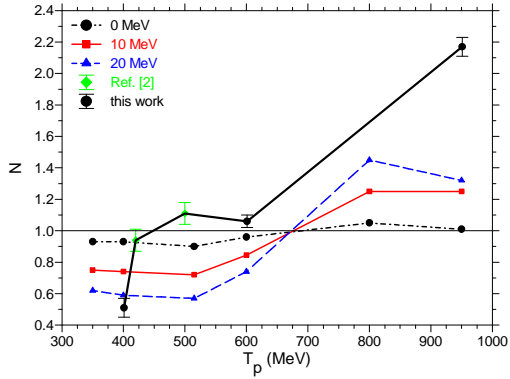}
\caption{(Color online) Normalization factor $N$ of Eq.~(\ref{equ:d_pn}), evaluated in
the present pion production model with the tensor force included, for
the indicated energies $\varepsilon$ in the $pn$ system. Also shown
are the values extracted from the experimental data at
$\varepsilon\approx 10$~MeV, as presented in Table~\ref{Tab:Norm}. }
\label{Fig:without}
\end{center}
\end{figure}
%
%

We now compare the results of the dynamical model with the
F\"{a}ldt-Wilkin FSI theorem of Eq.~(\ref{equ:d_pn}) by evaluating
the ratio $R_{pn/d}$ and hence the normalization factor $N$. The
values with the tensor force included are shown for a range of beam
energies in Fig.~\ref{Fig:without}. Below 650~MeV, the ratio is less
than unity while for higher energies it is above, and this feature
remains when the tensor force is omitted. It is remarkable is that,
within the dynamical model used here, the FSI theorem is also
satisfied to a high accuracy for mutually consistent wave functions
even when the tensor force is switched on. It is important to note
that the beam energy dependence of the calculations reflects
qualitatively the behavior of the data, which are also shown in
Fig.~\ref{Fig:without} for $\varepsilon=10$ MeV. Since this is very
similar whether the tensor force is included or not, it suggests that
the coupling between the $S$ and $D$ states is not primarily
responsible for the energy dependence of $N$ and that this probably
arises from the long range components in the production
operator~\cite{Niskanen78,Niskanen94}.

In summary we have measured the forward cross section for the
$pp\to\pi^+X$ reaction at two energies, one corresponding to the
center of the $\Delta$ resonance and one below. The high resolution
achieved allowed a clear separation of the two channels, $X=d$ and
$X=pn$. Comparison of the production of the two channels using the
F\"{a}ldt-Wilkin theorem~\cite{FW96} shows the necessity for the
introduction of normalization factors $N$. These factors together,
with one derived earlier~\cite{Abdel-Bary05a} and two obtained by
extrapolating TRIUMF data~\cite{Pleydon99} to zero degrees, follow a
smooth dependence as function of the beam energy. The comparisons
were made for $\varepsilon\approx 10$~MeV so that some deviations
might arise from $P$-wave $pn$ pairs. However, these contribute
incoherently and would not change the picture qualitatively.

To check the suggestion that the energy dependence of $N$ stems from
an interference between the $S$ and $D$ state in the continuum, model
calculations were performed for both $\pi^+d$ and $\pi^+pn$ final
states with and without the tensor force. For the lower bombarding
energies both types of calculation reproduce the data in both
channels. For the highest energy only that with tensor force accounts
for the continuum part of the data, though the deuteron pole is still
overestimated. The theoretical results of Fig.~\ref{Fig:without} give
a semi-quantitative understanding of the energy dependence of the FSI
normalization factor $N$ shown in Table~\ref{Tab:Norm}. With or
without the tensor force, the curves cross unity around $T_p\approx
650$~MeV, which corresponds to the maximum of the $pp\to \pi^+d$
excitation function and hence to the center of $\Delta(1232)$
production. The tensor force seems therefore not too be the dominant
effect in the energy dependence of $N$.

Although the F\"{a}ldt-Wilkin theorem was originally proven only for
$S$-waves, it was unexpected that the dynamical calculations with the
tensor force also satisfy the theorem when extrapolated to the
deuteron pole. However, even if the theorem is valid at the pole
itself, when it is used as an approximation to the $pn/d$ ratio in
the continuum it is seen that this can lead to errors, especially
above the $\Delta$ resonance. These deviations, shown as a function
of $\varepsilon$ in Fig.~\ref{Fig:without}, may arise from the long
range part of the pion production operator associated with the
on-shell intermediate pions. This is supported by the fact that $N$
changes from below to above one at an energy which corresponds to the
$\Delta$ excitation but also to the two pion threshold.

We are grateful to W.R.~Falk for making the data of
Ref.~\cite{Pleydon99} available to us. The quality of the beam
necessary for the success of this work is due mainly to the efforts
of the COSY operator crew. Support by the European Union under the
FP6 ``Structuring the European Research Area'', contract No.\
RII3-CT-2004-506078, from the Indo-German bilateral agreement
(Internationales B\"{u}ro des BMBF, IND 05/02), from the Research
Center J\"{u}lich (FFE), Academy of Finland (54038), DAAD 313-PPP-SF
and from GAS Slovakia (1/4010/07), SCSR Poland are all gratefully
acknowledged.

%
%

\end{document}